\documentclass[letterpaper]{aa} % for the letters 
\usepackage{graphicx}
\usepackage{txfonts}
\usepackage{natbib}
\begin{document}
   \title{Photometry of two DQ white dwarfs - search for spots}

   \subtitle{}

   \author{T. Vornanen
          \inst{1}
          \and
          A. Berdyugin\inst{1}
          }

   \institute{Department of Physics and Astronomy, University of Turku,
              Vaisalantie 20, FI-21500 Piikkio\\
              \email{tommi.vornanen@utu.fi}
             }

   \date{}

% \abstract{}{}{}{}{} 
% 5 {} token are mandatory
 
  \abstract
   {}
   {The intensity profiles of the C$_2$ Swan bands in cool DQ white dwarfs cannot be adequately fitted with models that otherwise succesfully reproduce spectral features of the molecule CH in these stars. Initial modelling showed that a two{-}component atmosphere in the style of a spot { might be able to} solve the problem.}
   {We photometrically observed the two cool DQ white dwarfs GJ1117 and EGGR78 to search for variability caused by stellar spots.}
   {We have not found any such variability, but we estimate the effects of hypothetical spots on lightcurves. We also estimate detection probabilities for spots in different configurations. Alternative explanations of the problem are needed and briefly discussed. }
   {}

   \keywords{ white dwarfs -- 
              stars: atmosphere --
              stars: spots
               }

   \maketitle 
%
%________________________________________________________________

\section{Introduction}

%White dwarfs (WDs) are the end product of the evolution of low mass stars. WDs are classified according to optical spectral features to subclasses denoted by DA (hydrogen), DB (neutral helium), DC (featureless), DZ (metals), DQ (carbon) and so on, as well as the combinations of these classes (DAB, for example). 

Magnetism in { white dwarfs (WDs)} is usually studied through Zeeman splitting of atomic spectral lines. This method has proven to be very useful, and with the sensitivity added by spectropolarimetry it is very effective in detecting magnetic fields. {Spectra of} cool DQ WDs do not show any atomic lines, { therefore} an alternative method is required. Fortunately, the carbon molecules present in { most of these stars} are sensitive to magnetic fields and they can be studied using spectropolarimetry.

%For the past several years we have been searching for magnetic fields in cool DQ WDs with carbon molecular bands using spectropolarimetry. Ever since the first time a DQ WD (\object{G99-37}) was found to be magnetic \citep{ang74}, no other such object has been found. Not until 2010, that is, when we published the discovery of a second cool DQ, \object{GJ841B} showing { signatures} of a magnetic field \citep{vor10}. This southern WD was found to be a close relative of G99-37. It has a similar temperature and a magnetic field of the same order of magnitude, albeit slightly weaker. 

{ During the past several years w}e have used circular spectropolarimetry to observe known DQ WDs to look for polarization signals from C$_2$ and CH molecules that can be found in their atmospheres. We have used two different telescopes, the Nordic Optical Telescope (NOT) { on} the Canary Islands and the ESO VLT (Cerro Paranal, Chile). So far, we have observed 12 objects and found one of them (GJ841B) to be magnetic \citep{vor10}. The rest of the stars do not show any polarization signal at the { noise} level of our observations, i.e. 0.5 \% { in Stokes V/I} for the NOT observations and 0.2 \% for the VLT observations.

\begin{figure}
%\epsscale{1.0}
%\plotone{EGGR78csC2_B_final.jpg}
\resizebox{\columnwidth}{!}{\includegraphics{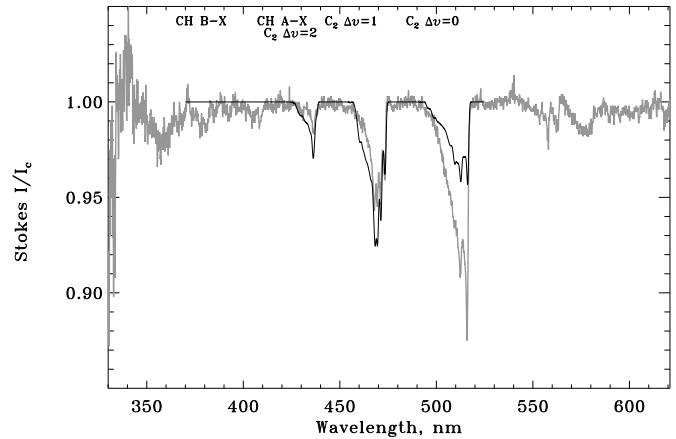}}
\caption{Model fit from \citet{ber05} applied to GJ893, one of the non-magnetic objects in our survey. Transition designations are
given above the spectrum. The discrepancy between the depths of different Swan bands is obvious. The temperature of the model is 6000K.} 
\label{fig:gj893}
\end{figure}

In \citet{vor10} we showed that the model presented in \citet{ber05, ber07} works very well for the CH absorption bands in GJ841B and were also successful in modelling the CH and C$_2$ blend at 430 nm visible in both G99-37 and GJ841B. But while we modelled the intensity profiles of the purely C$_2$ Swan bands, we were unable to achieve good fits. Problems with fitting Swan bands in DQ WDs have been reported before (for example, \citealt{duf05} have { shown} that the $\Delta v=+2$ bands are consistently too strong in their model fits). Our model shows the same problem (as seen { in} Fig. \ref{fig:gj893}), but in addition to that the $\Delta v=0$ band appears too weak. Hints of a too weak $\Delta v=0$ band can also be seen in \citet{duf05}. These problems persisted for all non-magnetic DQ WDs in our survey, as long as we used a single temperature model. 

We started searching for a solution to this problem by combining two models in { the} manner of a photosphere with a spot to { obtain}  reasonable fits to the spectra. This led to somewhat strange results. The best{ -}fitting models usually had a small spot with $T=8000-10000$ K and a photosphere of $T=2000$ K. This would suggest a cool disk or an envelope of gas or dust around the WD with a hole in it through which the atmosphere would be seen. Although dust disks have been found around hot WDs and the central objects of planetary nebulae \citep{bil11}, this does not sound like a very probable scenario for our cool objects. Instead, { a cool spot has} been found before on a weakly (70 kG) magnetic WD \citep{bri05}.

To exclude spots or envelopes with holes as the reason for the observed properties in these WDs, we started a photometric observing { programme} with a small remotely controlled telescope on { the} Canary Islands. Since WDs are rotating just like any other star, the movement of the spot on the visible hemisphere would cause some photometric variability. We report the results of our photometric observations here and discuss what we can deduce from them. At the end of the paper we discuss other possible reasons for troubles with modelling.

\section{Observations}

For our observations we used the 35cm Schmidt-Cassegrain Celestron telescope attached to the side of the 60cm KVA (Kungliga Vetenskapsakademien) telescope in Observatorio del Roque de Los Muchachos (ORM) on the island of La Palma in the Canary Islands. The telescope is remotely operated by researchers at Tuorla Observatory of the University of Turku (Finland). The small telescope has filters for BVRI photometry and is equipped with an Apogee Santa Barbara 47P CCD camera. The larger main telescope is used for linear polarization observations with the Dipol-polarimeter \citep{pii05}. The two telescopes are pointed to the same target and can be used simultaneously, if desired.

\begin{table}
\caption{Observing log}             % title of Table
\label{table1}      % is used to refer this table in the text
\centering                          % used for centering table
\begin{tabular}{l c c c}        % centered columns (4 columns)
\hline\hline                 % inserts double horizontal lines
Object & Date & N(B) & N(R) \\    % table heading 
\hline                        % inserts single horizontal line
EGGR78 &   2011-05-03 &  26  & \\
       &   2011-05-04 &  13  &  14 \\
       &   2011-05-05 &  16  & \\
       &   2011-05-09 &  23  & \\
       &   2011-05-10 &  23  & \\
       &   2011-05-12 &   3  & \\
       &   2011-05-13 &  23  & \\ 
       &   2011-05-15 &  25  & \\
       &   2011-05-16 &  23  & \\
       &   2011-05-17 &  23  & \\
       &   2011-05-20 &  17  & \\
       &   2011-05-22 &  19  & \\
       &   2011-05-23 &  20  & \\
       &   2011-05-24 &  15  & \\
       &   2011-05-29 &  16  & \\
       &   2011-05-30 &  16  & \\
       &   2011-05-31 &  12  & \\
       &   2011-06-02 &  16  & \\
       &              &      & \\
GJ1117 &   2011-03-22 &  7  &  8 \\
       &   2011-03-23 &  24  & \\
       &   2011-03-24 &  11  & \\
       &   2011-03-25 &  13  & \\
       &   2011-03-28 &  10  & \\
       &   2011-03-29 &  16  & \\
       &   2011-03-31 &  27  & \\
\hline                                   %inserts single line
\end{tabular}
\end{table}

\begin{figure}
%\epsscale{1.0}
%\plotone{EGGR78csC2_B_final.jpg}
\resizebox{\columnwidth}{!}{\includegraphics{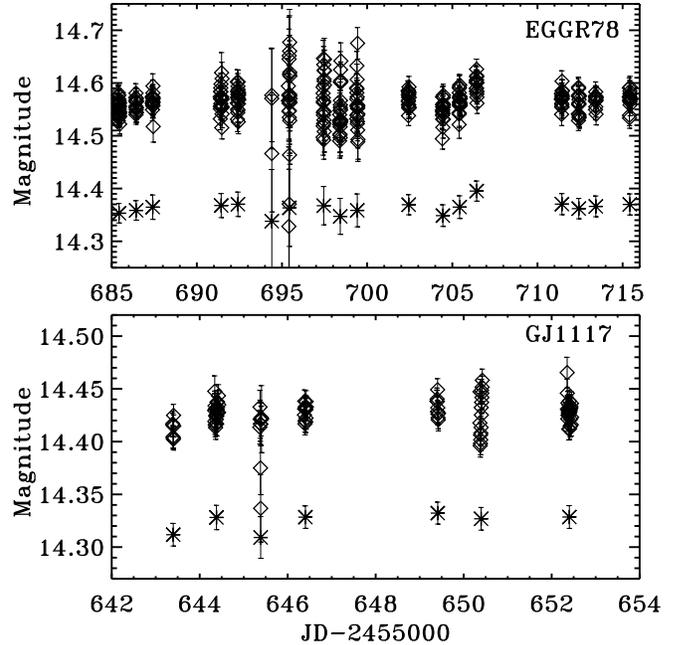}}
\caption{{ Lightcurves for EGGR78 and GJ1117 together with nightly averages (displaced for clarity).}} 
\label{fig:gj1117B}
\end{figure}

We took data for \object{GJ1117} during seven nights spread over ten nights and for \object{EGGR78} during 17 nights spread over one month. We took exposures in both {\it B} and {\it R} bands on all nights. The dates of the observations and number of exposures for each night are listed in Table \ref{table1} (available electronically only). Exposure times were 180 seconds for {\it B}-band and 100 seconds for {\it R}-band. We performed differential photometry comparing the objects to three comparison stars in the field of view. Since there is no information on the comparison stars apart from magnitudes in a few different bands (not {\it UBVR}), we also had to compare the comparison stars to each other to find out if any of them are variable. None of them were found to be variable within the errors. Figure \ref{fig:gj1117B} shows the { lightcurves for EGGR78 and GJ1117} in {\it B}-band together with nightly averages compared to another star in the field of view. 
%Figure \ref{fig:gj1117B} shows a similar plot for GJ1117 in {\it B}-band.
Errors of the individual measurements are from photon noise and the error on nightly averages are standard deviations of all measurements for the night in question. The magnitudes are arbitrary.

%\begin{figure}
%%\epsscale{1.0}
%%\plotone{EGGR78csC2_B_final.jpg}
%\resizebox{\columnwidth}{!}{\includegraphics{EGGR78C2B_dp_ave.eps}}
%\caption{Light curve for EGGR78 together with nightly averages (displaced for clarity).} 
%\label{fig:eggr78B}
%\end{figure}

%\begin{figure}
%%\epsscale{1.0}
%%\plotone{EGGR78csC2_B_final.jpg}
%\resizebox{\columnwidth}{!}{\includegraphics{GJ1117C1B_dp_ave.eps}}
%\caption{Light curve for GJ1117 together with nightly averages (displaced for clarity).} 
%\label{fig:gj1117B}
%\end{figure}

The KVA telescopes are mostly used for optical monitoring of blazars in connection with the MAGIC Telescopes also located at ORM \citep[See][for example]{alb06}. The stability limit of the telescope is 15 mmag. { As can be seen from Fig. \ref{fig:gj1117B}, there is no real variability in EGGR78 or GJ1117 in {\it B} or in {\it R} band (not shown here)}. 

\begin{figure}
%\epsscale{1.0}
%\plotone{EGGR78csC2_B_final.jpg}
\resizebox{\columnwidth}{!}{\includegraphics{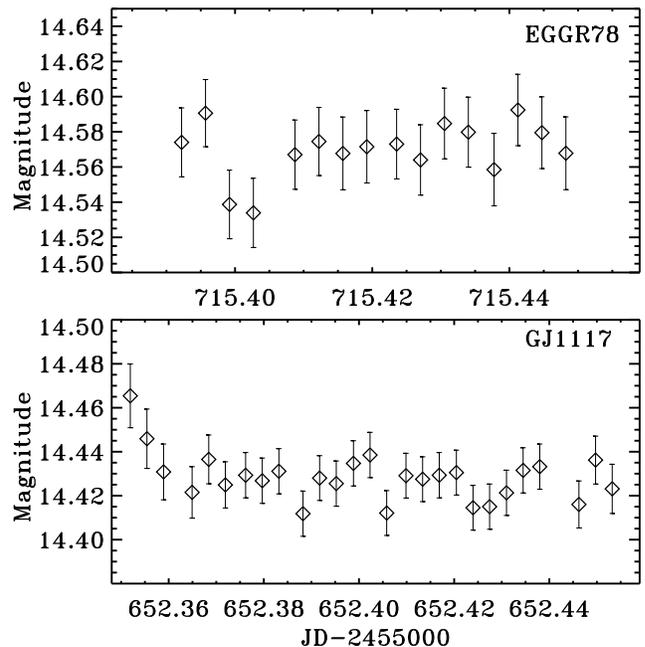}}
\caption{{ Lightcurves for EGGR78 and GJ1117 from the last night of observations (June 2nd and March 31st, 2011, respectively).}} 
\label{fig:eggr78B_last}
\end{figure}

{ Figure \ref{fig:eggr78B_last} (available electronically only) shows as an example a lightcurve from the last night of observations for each star.} No intra-night variability is evident in either object within errors.

%\begin{figure}
%%\epsscale{1.0}
%%\plotone{EGGR78csC2_B_final.jpg}
%\resizebox{\columnwidth}{!}{\includegraphics{EGGR78C2B_dp_lastnight_ave.eps}}
%\caption{Light curve for EGGR78 from the last night of observations (June 2nd, 2011).} 
%\label{fig:eggr78B_last}
%\end{figure}

%\begin{figure}
%%\epsscale{1.0}
%%\plotone{EGGR78csC2_B_final.jpg}
%\resizebox{\columnwidth}{!}{\includegraphics{GJ1117C1B_dp_lastnight_ave.eps}}
%\caption{Light curve for GJ1117 from the last night of observations (March 31st, 2011).} 
%\label{fig:gj1117B_last}
%\end{figure}

\begin{figure}
%\epsscale{1.0}
%\plotone{EGGR78csC2_B_final.jpg}
\resizebox{\columnwidth}{!}{\includegraphics{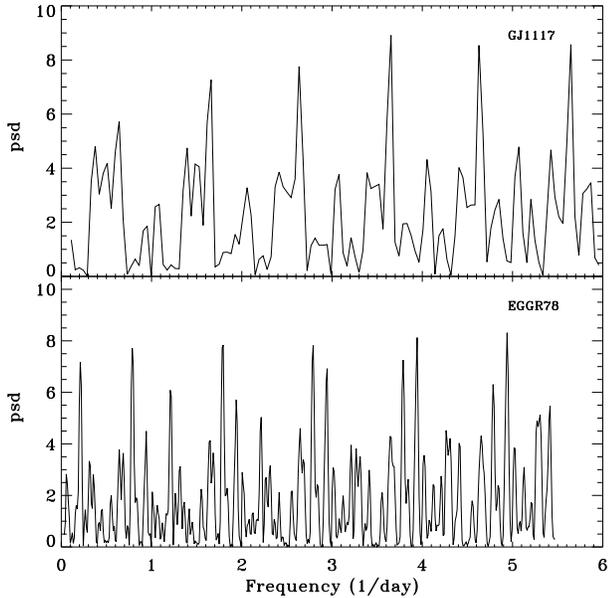}}
\caption{Periodograms for GJ1117 (top) and EGGR78 (bottom) show no signs of periodic variability.} 
\label{fig:periods}
\end{figure}

{ The periodograms (Figs. \ref{fig:periods}) are also} devoid of any signs of variability apart from the roughly 1 day period from the observing times.

%%%%%%%%%%%%%%%%%%%%%%%%%%%%%%%%%%%%%%%%%%%%%%%%%%%%%%
\section{Results}
%%%%%%%%%%%%%%%%%%%%%%%%%%%%%%%%%%%%%%%%%%%%%%%%%%%%%%

Obtained lightcurves appear to be flat within the errors and neither inspection by eye, nor a Lomb--Scargle periodogram \citep{lom76,sca82} can detect any kind of variability in our data set. 

Since we did not find any { signatures} of variability in our photometry, we { considered} carefully to { which} kind of variations we are sensitive to and { which} we might be missing. We have two sources of uncertainty here. We might miss variability { weaker} than 15 mmag { because of} the limiting accuracy of our { small} telescope, and the timings of our observations also limit our sensitivity to certain rotational periods. 

To estimate the amplitude variations we could expect in the lightcurves, we used the model discussed earlier to calculate a lightcurve when a spot of higher C$_2$ concentration and a temperature different from the atmospheric temperature crosses the visible hemisphere of the white dwarf. { For simplicity}, we assumed the spot to be a circular, flat disc. { The angle between the normal of this disc and the line of sight is given by
\begin{equation}
\cos{\alpha}=\cos(\,\beta\,)\sin(\,i\,)\cos(\,\phi\,)+\sin(\,\beta\,)\cos(\,i\,),
\end{equation}
where $\beta$ is the spot latitude, $i$ is the inclination of the rotation axis from the line of sight, and $\phi$ is the rotational phase ($\phi$=0 towards the line of sight) \citep{bri05}. Conveniently, this is also the size of the spot projected { onto} the plane of the sky. We also ignored limb{ -}darkening effects for simplicity.} The model produces { normalised} intensity spectra, { therefore} we had to transform our observed lightcurve from magnitudes to fluxes and then { normalise} them to the average value of the flux to { obtain} comparable results from observations and modelling.

\begin{figure}
%\epsscale{1.0}
%\plotone{EGGR78csC2_B_final.jpg}
\resizebox{\columnwidth}{!}{\includegraphics{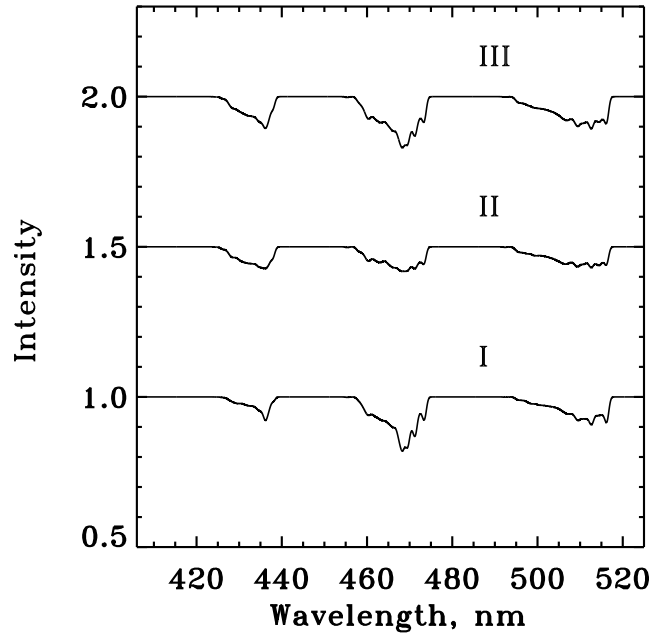}}
\caption{Spot models: (I) homogenous atmosphere of 8000 K, (II) spot of 8000 K in an atmosphere of 2000 K, and (III) spot of 6000 K in an atmosphere of 8000 K.} 
\label{fig:spotmodels}
\end{figure}

We started by calculating the lightcurve from the spot/atmosphere combinations suggested by fitting the observed Stokes I spectra of the two WDs. The temperatures in these two models are 2000 K for the atmosphere and 8000 K and 10000 K for the spots in EGGR78 and GJ1117, respectively, with spot sizes of 1 \% and 10 \%. In both stars, modelling suggests that there is a low concentration of C$_2$ in the atmosphere in general and the spot has a much higher amount of C$_2$. Figure \ref{fig:spotmodels} shows three examples of model spectra: (I) homogenous atmosphere at 8000 K, (II) spot of 8000 K in an atmosphere of 2000 K, and (III) a more realistic 6000 K spot in an atmosphere of 8000 K. The differences between the spectra are very small, especially in cases I and III.

\begin{figure}
%\epsscale{1.0}
%\plotone{EGGR78csC2_B_final.jpg}
\resizebox{\columnwidth}{!}{\includegraphics{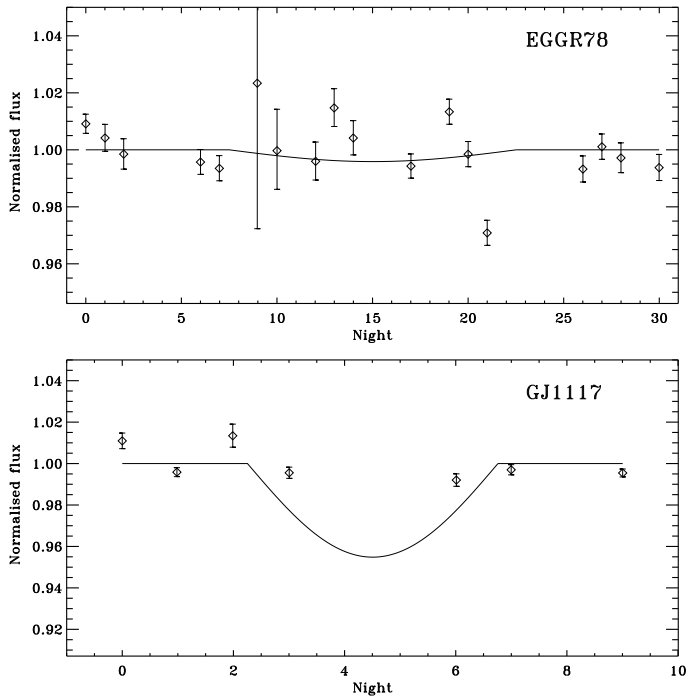}}
\caption{Observed lightcurves (diamonds) of EGGR78 (top) and GJ1117 (bottom) together with model lightcurves (solid line).} 
\label{fig:model_lightcurves}
\end{figure}

Since we do not know the rotational period of the white dwarfs, we used the time span of the photometric observations as the periods for illustration purposes. Results are shown in Fig. \ref{fig:model_lightcurves}.

It is evident { from Fig. \ref{fig:model_lightcurves}} that the spot model that fits the spectrum of EGGR78 will not produce a { sufficiently deep} dip in the lightcurve for us to detect it. It would take a size of nearly 10 \% of the visible hemisphere for the spot to be detectable. This { is because} the C$_2$ Swan bands on EGGR78 are very weak and the filter sensitivity makes the variability of the absorption bands even harder to detect, since the strongest band between 480 and 520 nm is not included in the $B$ filter. Variability on GJ1117 would clearly have been found if it were present. 

{ The model lightcurves in Fig. \ref{fig:model_lightcurves} have the most ideal spot configuration for detection: namely, a spot on the equator of the star. When the spot is moved towards the pole, the drop in the lightcurve { becomes} shallower as the projected area of the spot, and its variation, becomes smaller. The same happens if the rotational axis is tilted from the plane of the sky towards the observer. This also changes the duration of the dip because the spot spends a different amount of time behind the star. Finally, when the spot latitude $\beta$ is { higher} than the inclination of the rotational axis $i$, the lightcurve assumes a true $sine$-form { because} the spot remains visible at all times but its projected size varies.}

{ To estimate our chances of detecting a spot depending on the system geometry, we calculated a lightcurve for all angles of spot latitude ($-90^\circ \ldots  90^\circ$) and inclination ($0^\circ \ldots  90^\circ$) in one degree intervals. If the difference between maximum and minimum values of brightness on the lightcurve was larger than 0.03, we counted it as detected. We also gave weights to the chances of detecting a spot for each latitude/inclination combination based on cos($\beta-i$) to account for the fact that the points appear to be more { densely} distributed away from the line of sight towards the edges of the stellar disk. In this way we estimated the chances of missing a spot on both stars. Using a spot size of 10 \% we calculated that we would have had { about a} 40 \% chance of missing a spot in both our WDs using either model II or III.}

%Using model (III) and a size of 10 \% the chances of missing a spot due to geometrical reasons are  76 \% for EGGR78 and 60 \% for GJ1117. The chance of missing a spot in both stars is then the product of these, i.e. 46 \%. Using model (II) the percentages were 67 \% for EGGR78, 62 \% for GJ1117 and 42 \% for missing both.}

There is, of course, the chance that we { did not} see variations in our photometric observations because the spot was behind the star during the observations. In this case, we might see some C$_2$ that has spread out over the entire star, and missed the spot that is needed to model the spectra taken a year earlier. With { the correct} values of temperature and carbon abundance the spot can just alter the shape of the molecular absorption bands without changing their depths. The variations would be very difficult to pick up for the untrained eye { by just} looking at the spectra. Moreover, no one has tried monitoring these kinds of stars spectroscopically for a longer period to see if they are variable or not.

We should also consider the effect of a very long rotational period { because} our photometric observations as well as the spectropolarimetric observations might have been taken in the deepest part of the brightness drop. The { normalisation} level of the model lightcurve is determined by our spectropolarimetric observations, which were taken a year before the photometric observations, { therefore} a long rotational period could mean that all our observations have been { performed} during a minimum. For example, a 3000{ -}day period would mean that both our spectropolarimetric and photometric observations could have been made in a low state.

\section{Conclusions}
%%%%%%%%%%%%%%%%%%%%%%%%%%%%%%%%%%%%%%%%%%%%%%%%%%%%%%

We monitored two cool DQ WDs to search for photometric variability as a { signature} of a stellar spot. We did not find { any} such variability. The accuracy and timing of our observations do not allow us to detect very long rotational periods (of { about} a year or longer), very small spots, or some spot configurations. Previous studies of WDs have come to the conclusion that the rotation periods usually { last} from hours to decades \citep[][and references therein]{cha09}.
% A few magnetic WDs have been found to have polarization features that are stable over a time span of a decade leading to rotational periods longer than 100 years \citep{sch91}. 
Some magnetic WDs have indeed been found to be slow rotators with periods of about 100 years \citep{schnor91, ber99}. Since we did not detect any intra-night variability, the shorter end of the period distribution can be ruled out. Based on these arguments, we can assume that GJ1117 and EGGR78 still might have very long rotational periods. 

%Unfortunately most of the cool DQ WDs do not have any atomic spectral lines that could be used to determine the rotational velocity using high resolution spectroscopy. GJ1117 is an exception in this sense because it shows some lines of neutral carbon. GJ1117 is, however, so faint (M$_V$=15.2) that obtaining high resolution spectra of the target with high enough accuracy is impossible with current instruments, especially since there are not that many carbon lines in its spectrum and the rotational velocity might be of the order $10^{-2}$m/s assuming a rotational period of a 100 years. 

Although we { cannot} rule out spots on these WDs completely, we consider the values given by our model for the spot sizes and temperatures { to be} very { unrealistic}. But if the spot model is not the solution to the discrepancy in intensity in the carbon molecular bands, { what} better hypotheses { could be found}? We think the answer may lie in the  oscillator strengths of carbon molecules { that are}  embedded in our model. These parameters determine { the strengths of} the individual absorption lines within the molecular bands. By modifying these values slightly, we hope to achieve consistent fits to all molecular bands without invoking a spot model. The current values { were} determined in laboratory conditions, but still contain significant error limits.

If modifying the oscillator strengths will not work, we have to go deeper into the physics of the problem and investigate { in which way} the structure, and therefore, the properties of a C$_2$ molecule is different { in} a white dwarf from conditions in { the} laboratory where the molecule { was} studied. Or maybe our treatment of the WD atmosphere is { incorrect} and we have to consider its properties more thoroughly.

%Why do we then have to use a spot model to get reasonable fits to our spectropolarimetric data? We believe the answer lies in our model, especially in the oscillator strengths. These parameters determine how strong the individual absorption lines within the molecular bands are. By modifying these values slightly we hope to achieve consistent fits to all of the molecular bands without invoking a spot model.

\begin{acknowledgements}
We would like to thank the referee for helpful comments. They improved this article immensely.

T.V. would like to thank the Finnish Graduate School for Astronomy and Space Physics for financial support that has made this study possible. The authors would also like to express their gratitude { to} Kari Nilsson for his help in using the Diffphot photometry data reduction package that he has created, which we used to reduce the photometric data for this paper.
\end{acknowledgements}

\bibliographystyle{aa}

\bibliography{/home/ttvorn/Desktop/omat_paperit/References/bibliography}

%\begin{thebibliography}{}

%\end{thebibliography}

\end{document}